\documentclass[10pt,twocolumn]{revtex4}
\usepackage{amssymb}
\usepackage{latexsym}
\usepackage{epsfig}
\usepackage{color}
\begin{document}
\title{Thermodynamic analysis of gravitational field equations in Lyra manifold}
\author{H. Moradpour\footnote{hn.moradpour@gmail.com}, N. Sadeghnezhad\footnote{nsadegh@riaam.ac.ir}, S. Ghaffari\footnote{ghaffarishayesteh@gmail.com}, A. Jahan\footnote{jahan@riaam.ac.ir}}
\address{Research Institute for Astronomy and Astrophysics of Maragha (RIAAM), P.O. Box 55134-441, Maragha, Iran}

\begin{abstract}
Considering the Einstein field equations in Lyra manifold, and
applying the unified first law of thermodynamics as well as the
Clausius relation to the apparent horizon of FRW universe, we find
the entropy of apparent horizon in Lyra manifold. In addition, the
validity of second law of thermodynamics and its generalized form
are also studied. Finally, we use the first law of thermodynamics
in order to find the horizon entropy of static spherically
symmetric spacetimes. Some results of considering (anti)de-Sitter
and Schwarzschild metrics have also been addressed.
\end{abstract}
\maketitle

\section{Introduction}

Inasmuch as some observational data are not suitably described by
the general relativity theory, physicists try to get gravitational
theories in more agreement with nature
\cite{Rev3,Rev1,mod,meeq,lobo,rep1}. In one of these approaches,
it has been shown that the accelerating universe may originate
from the breaking of conformal symmetry \cite{rep1}. The
Riemannian geometry is the backbone of most of these theories, and
it has been shown that the system horizon (as the boundary of
system) has entropy, and indeed, there is a deep connection
between the thermodynamics laws and the gravitational field
equations
\cite{haw,GSL,GSL0,jacob,jacob1,mis1,ref0,ref1,ref2,pad,pad1,pad2,pad3,j1,j2,Cai2,CaiKim,j4,j5,j6,j7,j8,j9,j10,j11,j12,j13,plb,ms}.
This mutual relation between gravity and thermodynamics may help
us in providing a description for the primary inflationary era as
well as the current accelerating phase of universe \cite{mra}. In
addition, due to such relation, geometry and the universe material
content may approach a thermodynamic equilibrium \cite{pavonf}.

In fact, the dark energy candidates, as the source of the current
phase of universe, may modify the horizon entropy and thus its
thermodynamics in both gravitational and cosmological setups
\cite{cana,cana1,mmg,em,mitra,mitra1,mitra2,mitra3,mms,md}.
Moreover, it seems that such modifications to the thermodynamics
of system are in line with probable non-extensive thermodynamic
properties of spacetime and the current universe
\cite{salis,ijtp,raf1}.

A generalization of the Riemannian manifold has been introduced by
Lyra \cite{L1} which modifies the Einstein field equations
\cite{sen}. Changing Reimannian manifold by the Lyra geometry, one
can also find a new scalar-tensor gravity in which both the tensor
and scalar fields have geometrical interpretations \cite{seng}.
Various aspects of the Einstein field equations in the Lyra
manifold have been studied
\cite{dv1,dv2,lrev,D1,s1,s2,hara,lyraobs,lkur,sepangi,dwhl}. Using
the Einstein field equations in Lyra manifold, Karade et al., have
studied thermodynamic equilibrium of a spherical gravitating fluid
\cite{lyrather}.

Here, using the Einstein field equations in Lyra manifold, we are going to study the mutual relation between the field
equations and the thermodynamics laws. Moreover, we are interested
in obtaining the effects of Lyra generalization of Reimannian
manifold on some thermodynamic properties of gravitating systems.
In order to achieve this goal, applying the thermodynamics laws to
the apparent horizon of FRW universe, we find an expression for
the horizon entropy in Sec.~($\textmd{II}$). In addition, the
validity of second and generalized second laws of thermodynamics
will also be investigated in the next section. In
Sec.~($\textmd{III}$), we use the Einstein field equations in Lyra
manifold as well as the unified first law of thermodynamics in
order to obtain the generalization of the Misner-Sharp mass
\cite{mis} in this theory. Moreover, we use the obtained mass
relation, field equations and the first law of thermodynamics to
find the entropy of static spherically symmetric horizons. A
summary is presented in section~($\textmd{IV}$). For the sake of
simplicity, we consider $8\pi G=c=\hbar=1$ in our study.

\section{Thermodynamics of the apparent horizon in Lyra manifold}

In the Lyra manifold, the Einstein field equations are written as

\begin{eqnarray}\label{Ein}
G_{\mu\nu}+\frac{3}{2}\phi_\mu\phi_\nu-\frac{3}{4}\phi_\alpha\phi^\alpha
g_{\mu\nu}=T_{\mu\nu},
\end{eqnarray}

\noindent where $\phi_\nu=g_{\mu\nu}\phi^\mu$ denotes the
displacement vector field of Lyra manifold, and we set Einstein
gravitational coupling constant to one \cite{L1,D1}. As it is
obvious, due to the geometrical field $\phi_\mu$, the Einstein
gravitational field equations in Lyra and Riemannian manifolds
differ from each other. Here, we are interested in investigating
the effects of the displacement vector field on the apparent
horizon entropy of the FRW universe, and thus its thermodynamics.

In the cosmological setup, there are two common cases for the
$\phi_\nu$ vector including $\phi_\nu=(\beta(t),0,0,0)$ and
$\phi_\nu=(\alpha,0,0,0)$, where $\alpha$ is constant, which help us
in describing the current accelerating universe \cite{dv1,dv2}.
Here, we focus on the general case of $\phi_\nu=(\beta(t),0,0,0)$,
and find the entropy of apparent horizon.

The FRW universe is described by

\begin{eqnarray}
ds^2=-dt^{2}+a^{2}\left( t\right) \left[ \frac{dr^{2}}{1-kr^{2}}%
+r^{2}d\Omega ^{2}\right],
\end{eqnarray}

where $a(t)$ is scale factor. In addition, $k=1,0,-1$ denotes the
curvature parameter corresponding to closed, flat and open
universes, respectively. Therefore,
$\phi_\nu\phi^\nu=\phi_0\phi^0=-\beta^2(t)$, and if universe is
filled by an energy-momentum source as
$T_\mu^\nu=diag(-\rho,p,p,p)$, then by defining the Hubble
parameter as $H\equiv\frac{\dot{a}}{a}$, one can reach at

\begin{eqnarray}\label{Fri1}
\rho&=&3[H^2+\frac{k}{a^2}]+\frac{3}{4}\beta^2(t),\\
p&=&-3H^2-2\dot{H}-\frac{k}{a^2}+\frac{3}{4}\beta^2(t),\nonumber
\end{eqnarray}

\noindent for the Friedmann equations in the Lyra geometry. Here,
$\rho$ and $p$ are the energy density and pressure of the
isotropic fluid which supports the geometry, respectively.
Moreover, here and from here onwards, dot denotes the derivative
with respect to time. One can also use the Friedmann equations in
order to find

\begin{eqnarray}\label{Rey}
\dot{H}-\frac{k}{a^2}=-\frac{\rho+p}{2}+\frac{3}{4}\beta^2(t).
\end{eqnarray}

\noindent In addition, since the apparent horizon of FRW universe
is located at

\begin{eqnarray}\label{ah}
\tilde{r}_A=a(t)r_A=\frac{1}{\sqrt{H^2+\frac{\kappa}{a(t)^2}}},
\end{eqnarray}

\noindent we can rewrite Eq.~(\ref{Rey}) in the form of

\begin{eqnarray}\label{Rey1}
\rho+p=\frac{2\dot{\tilde{r}}_A}{H\tilde{r}^3_A}+\frac{6}{4}\beta^2(t).
\end{eqnarray}

In order to obtain continuity equation in our setup, one can
define an energy-momentum tensor corresponding to the displacement
vector field $\phi_\mu$ as

\begin{eqnarray}\label{enphi}
T_\mu^\nu(\phi)=\frac{3}{4}\phi_\alpha\phi^\alpha
\delta_\mu^\nu-\frac{3}{2}\phi_\mu\phi^\nu,
\end{eqnarray}

\noindent leading to

\begin{eqnarray}\label{enphi1}
T_\mu^\nu(\phi)=diag(-\rho(\phi),p(\phi),p(\phi),p(\phi)),
\end{eqnarray}

\noindent where $\rho(\phi)=p(\phi)=-\frac{3}{4}\beta^2(t)$ are
the energy density and isotropic pressure corresponding to the
displacement vector field of $\phi_\nu=(\beta(t),0,0,0)$. If we
define the state parameter corresponding to the Lyra displacement
vector field as $w_{\phi}\equiv\frac{p(\phi)}{\rho(\phi)}$, then
we have $w_{\phi}=1$ meaning that the displacement vector field
acts as a stiff matter. Moreover, since
$\rho(\phi)=-\frac{3}{4}\beta^2(t)$ is negative for real non-zero
values of $\beta(t)$, the weak energy condition is not respected
by a real non-zero $\phi_\mu$. In this manner, we have also
$p(\phi)<0$ for $\beta(t)\neq0$ meaning that the real Lyra
displacement vector field induces a negative pressure into the
background. This vector field may help us to build a model for an
accelerating universe \cite{dv1,hara} which can avoid the
singularity, entropy and horizon problems \cite{lrev}. It is also
worth to mention that for imaginary Lyra displacement vector
fields we have $\rho(\phi)=p(\phi)>0$ meaning that the weak energy
condition is satisfied. Therefore, since such fields do not lead
to negative pressure, as the basic property of the accelerating
universe, this kind of Lyra displacement vector field cannot be
used to describe the inflationary and current phases of universe.

Let us define the total energy-momentum tensor ($\Theta_\mu^\nu$)
as $\Theta_\mu^\nu\equiv
T_\mu^\nu+T_\mu^\nu(\phi)=diag(-\rho_e,p_e,p_e,p_e)$ in which
$\rho_e=\rho+\rho(\phi)$ and $p_e=p+p(\phi)$. In this manner, if
$\rho\geq-\rho(\phi)=\frac{3}{4}\beta^2(t)$, then the
$\rho_e\geq0$ condition is respected. Moreover, rewriting the
Einstein field equations in Lyra manifold as $G^{\
\nu}_{\mu}=\Theta^{\ \nu}_{\mu}$, one can easily obtain that the
$\rho_e\geq0$ condition is satisfied whenever the Hubble parameter
meets the $H^2\geq0$ condition, or equally $H$ should be a real
quantity.

The Bianchi identity ($G^{\ \nu}_{\mu\ ;\nu}=0$) implies $\Theta^{\
\nu}_{\mu\ ;\nu}=0$ which leads to

\begin{eqnarray}\label{cont1}
\dot{\rho}+3H(\rho+p)=\frac{3}{2}[\beta(t)\dot{\beta}(t)+3H\beta^2(t)],
\end{eqnarray}

\noindent for the continuity equation. This equation can also be
decomposed into

\begin{eqnarray}\label{cont2}
\dot{\rho}+3H(\rho+p)=0,
\end{eqnarray}

\noindent and

\begin{eqnarray}\label{cont3}
\dot{\beta}(t)+3H\beta(t)=0,
\end{eqnarray}

\noindent whenever there is no interaction between $\phi_\mu$ and
the energy-momentum source.

\subsection{The entropy of apparent horizon}

The Cai-Kim temperature of the apparent horizon and the projection
of the four-dimensional energy-momentum tensor $T^b_a$ on the
normal direction of the two-dimensional sphere with radius
$\tilde{r}$ are as follows

\begin{eqnarray}\label{aht}
T=\frac{1}{2\pi\tilde{r}_A},
\end{eqnarray}

\noindent and

\begin{eqnarray}\label{esv}
\psi_a = T^b_a\partial_b \tilde{r} + W\partial_a \tilde{r},
\end{eqnarray}

\noindent respectively \cite{CaiKim,Cai2,CaiKimt}. Here,
$W=\frac{\rho-p}{2}$ is the work density. Additionally, the energy
flux ($\delta Q^m$) crossing the system boundary (the apparent
horizon) is defined as $\delta Q^m \equiv A\psi_a dx^a$, and
therefore, simple calculations lead to

\begin{eqnarray}\label{ufl3}
dS_A=6\pi\tilde{r}_AVH(\rho+p)dt= 8\pi^2H\tilde{r}_A^4(\rho+p)dt,
\end{eqnarray}

\noindent where we used the $TdS_A\equiv-\delta Q^m$ (the Clausius relation) and $V=\frac{4\pi}{3}\tilde{r}_A^3$ (aerial volume) relations to obtain this
equation \cite{CaiKim,Cai2,CaiKimt,plb}. Now, inserting
Eq.~(\ref{Rey1}) into this equation and integrating the results,
one can easily show

\begin{eqnarray}\label{uflf}
S_A=S_{B}+12\pi^2\int\frac{H\beta^2(t)}{(H^2+\frac{k}{a^2})^2}dt,
\end{eqnarray}

\noindent where $S_B=2\pi A=8\pi^2\tilde{r}_A^2$ is the horizon
entropy in the Einstein framework (or equally the Bekenstain
entropy) if the Einstein field equations are written as
$G^{\mu\nu}=T^{\mu\nu}$. Indeed, since $8\pi G=\hbar=c=1$, we have
$2\pi=\frac{1}{4G}$ and thus $S_B=\frac{A}{4G}$ meaning that, at
the $\beta(t)\rightarrow0$ limit, the results of the Einstein
theory is covered. Therefore, the second term on the RHS of this
equation is the correction term to the horizon entropy due to the
displacement vector filed of $\phi_\nu=(\beta(t),0,0,0)$
considered in the Lyra geometry.

If one writes the Einstein field equations as $G^{\mu\nu}=8\pi
T^{\mu\nu}$, then simple calculations lead to
$S_A=\frac{S_{B}}{8\pi}=\frac{A}{4}$ for the horizon entropy.
Moreover, it has also been shown that if the Einstein field
equations are modified as $G^{\mu\nu}=8\pi\Theta^{\mu\nu}$, where
$\Theta^{\mu\nu}$ includes both the ordinary energy-momentum
tensor ($T^{\mu\nu}$) and correction terms ($\tau^{\mu\nu}$), then
one obtains

\begin{equation}\label{q}
S_A=\frac{A}{4}-8\pi^2\int \frac{H(\rho_c
+p_c)}{(H^2+\frac{k}{a^2})^2}dt,
\end{equation}

\noindent for the horizon entropy
\cite{em,mitra,mitra1,mitra2,mitra3,mms,md}. Here, $\rho_c$ and
$p_c$ denote the energy density and pressure components of
$\tau^{\mu\nu}$, respectively. It is useful to mention here that, irrespective of a mutual interaction between
various parts of $\Theta^{\mu\nu}$,
the above result is available \cite{mms,md}. Therefore, one must expect that replacing $\frac{A}{4}$ by $S_B$ in Eq.~(\ref{q}), we
should get Eq.~(\ref{uflf}). As we have previously shown, in our
case $\rho_c(\phi)=p_c(\phi)=-\frac{3}{4}\beta^2(t)$. Thus,
replacing $\frac{A}{4}$ by $S_B$ and inserting
$\rho_c(\phi)=p_c(\phi)=-\frac{3}{4}\beta^2(t)$ into
Eq.~(\ref{q}), one can easily reach Eq.~(\ref{uflf}) for the
horizon entropy.

Finally, it is worthwhile mentioning that since we have not used
Eqs.~(\ref{cont2}) and~(\ref{cont3}) to obtain Eq.~(\ref{uflf}),
this equation is valid irrespective of existence or absence of an
interaction between Lyra displacement vector field and the
energy-momentum source. This is in agreement with the properties of Eq.~(\ref{q}).

\subsection{The second and generalized second laws of thermodynamics}

From Eq.~(\ref{ufl3}), it is obvious that, independent of
$\phi_\mu$, the second law of thermodynamics, stated as
$\frac{dS_A}{dt}\geq0$ \cite{haw}, is met if the energy-momentum
tensor satisfies the $\rho+p\geq0$ condition. Therefore, if we
define the state parameter as $w=\frac{p}{\rho}$, then the
$\rho+p\geq0$ condition is covered for sources of positive energy
density ($\rho\geq0$) only if we have $w\geq-1$.

Moreover, the generalized second law of thermodynamics (GSLT)
states the total entropy of system ($S_T$), including the horizon
entropy ($S_A$) and the entropy of fluid ($S_m$) which supports
the geometry, should not decrease or equally it has to respect the
$\frac{dS_T}{dt}=\frac{dS_A}{dt}+\frac{dS_m}{dt}\geq0$ condition
\cite{GSL,GSL0}. In order to find the $dS_m$ term, we use the
first law of thermodynamics \cite{CALLEN}

\begin{equation}\label{Gibbs1}
T_mdS_m=dE+pdV,
\end{equation}

\noindent where $T_m$ is the temperature corresponding to the
supporter fluid. Inserting the $E=\rho V$ relation into
Eq.~(\ref{Gibbs1}), one finds

\begin{eqnarray}\label{entm}
\frac{dS_m}{dt}=2\pi\tilde{r}_A[(\rho+p)\dot{V}+V\dot{\rho}]
\end{eqnarray}

\noindent where we also used the $T_m=T=\frac{1}{2\pi\tilde{r}_A}$
relation \cite{pavonf}. Now, for non-interacting case, using
Eqs.~(\ref{cont2}),~(\ref{ufl3}) and the above result, we reach at

\begin{eqnarray}\label{GSLTn}
\frac{dS_T}{dt}=\frac{dS_A}{dt}+\frac{dS_m}{dt}=6\pi
V(\rho+p)\dot{\tilde{r}}_A,
\end{eqnarray}

\noindent for the total entropy changes. Therefore, for a
non-interacting expanding universe in which $\dot{\tilde{r}}_A>0$,
GSLT is satisfied if the dominant fluid respects the $\rho+p\geq0$
condition. It is also obvious that, for spacetimes in that
$\dot{H}=0$, we have
$\dot{\tilde{r}}_A=\frac{Hk}{a^2(\sqrt{H^2+\frac{k}{a^2}})^3}$
meaning that, only in the flat and closed universe, GSLT is
satisfied for a fluid respecting the $\rho+p\geq0$ condition. In
addition, for an open universe, GSLT is met if we have
$\rho+p\leq0$. Moreover, considering an interacting universe in
which Eq.~(\ref{cont2}) is not valid, we use the original form of
the continuity equation~(\ref{cont1}) and Follow the recipe which
led to Eq.~(\ref{GSLTn}) in order to get

\begin{eqnarray}\label{GSLTn1}
\frac{dS_T}{dt}=6\pi\tilde{r}_AV[\frac{\dot{\tilde{r}}_A(\rho+p)}{\tilde{r}_A}+\frac{\beta(t)}{2}(\dot{\beta}(t)+3H\beta(t))],
\end{eqnarray}

\noindent for the total entropy. Therefore, GSLT is valid whenever
we have
$\frac{\dot{\tilde{r}}_A(\rho+p)}{\tilde{r}_A}+\frac{\beta(t)}{2}(\dot{\beta}(t)+3H\beta(t))\geq0$.
It has been shown that a constant $\beta(t)$ may help us in
providing a description for the current acceleration phase of the
universe expansion and thus the cosmological constant \cite{dv1}.
Therefore, we now focus on the $\beta(t)=\beta=constant$ case,
leading to

\begin{eqnarray}\label{GSLTn2}
\frac{dS_T}{dt}=6\pi\tilde{r}_AV[\frac{\dot{\tilde{r}}_A(\rho+p)}{\tilde{r}_A}+\frac{3H\beta^2}{2}],
\end{eqnarray}

\noindent meaning that GSLT is met when we have
$\beta^2\geq\frac{2\tilde{r}_A^2(\rho+p)}{3}(\dot{H}-\frac{k}{a^2})$
yielding $\beta^2\geq0$ for a flat universe ($k=0$) of constant
Hubble parameter ($\dot{H}=0$). In obtaining the latter result, we
used the
$\dot{\tilde{r}}_A=H\frac{\frac{k}{a^2}-\dot{H}}{(\sqrt{H^2+\frac{k}{a^2}})^3}$
relation.

\section{Thermodynamics of the static horizon}

We consider a four-dimensional static, spherically symmetric
spacetime with a horizon located at $r_h$ and described by the
metric

\begin{equation}\label{metric}
ds^2=-f(r)dt^2+\frac{dr^2}{f(r)}+r^2d\Omega^2.
\end{equation}

Inserting metric~(\ref{metric}) in Eq.~(\ref{Ein}), the (tt) and
(rr) components of the Einstein tensors can be written as

\begin{equation}\label{Gtt}
G^0_{\ 0}=G^1_{\ 1}=\frac{1}{r^2}(rf^\prime(r)+f(r)-1),
\end{equation}

\noindent where prime denotes derivative with respect to $r$. We
consider a perfect fluid with $T_\mu^\nu=diag(-\rho,p,p,p)$ which
fills the spacetime. Therefore, since both the Einstein and
energy-momentum tensors are diagonal, only one component of the
Lyra displacement vector field may be non-zero and its other
components should be zero. We consider a timelike Lyra
displacement vector field as $\varphi_\nu=(\varphi,0,0,0)$, where
$\varphi$ can be a function of $r$ or a constant \cite{lyrather,s2}, and insert Eq.~(\ref{Gtt}) into
Eq.~(\ref{Ein}) to obtain \cite{lyrather}

\begin{equation}\label{rho}
\rho=\frac{1}{r^2}(1-rf^\prime(r)-f(r))-\frac{3}{2}\varphi_0\varphi^0+\frac{3}{4}\varphi_{\alpha}\varphi^{\alpha}.
\end{equation}

One can use Eq.~(\ref{esv}) as well as the $dE \equiv A\psi_a
dx^a$ relation in order to reach at \cite{ms}

\begin{equation}
dE=4\pi r^2\rho dr,
\end{equation}

\noindent which leads to

\begin{equation}\label{rho20}
dE=4\pi\Big(1-rf^\prime(r)-f(r)-\frac{3}{2}r^2\varphi_0\varphi^0+\frac{3}{4}r^2\varphi_{\alpha}\varphi^{\alpha}\Big)dr,
\end{equation}

\noindent and thus

\begin{equation}\label{rho2}
dE=4\pi\Big(1-\frac{d(rf(r))}{dr}+\frac{3r^2}{4f(r)}\varphi^2\Big)dr,
\end{equation}

\noindent where we also used the
$\varphi_{\alpha}\varphi^{\alpha}=\varphi_0\varphi^0=-\frac{1}{f(r)}\varphi^2$
relation to find the above result. We finally reach at

\begin{equation}\label{fe}
E=4\pi r(1-f(r))+3\pi\int\frac{\varphi^2r^2}{f(r)}dr,
\end{equation}

\noindent for the Misner-Sharp mass confined to radius $r$ in this
theory. Therefore, for total mass of a black hole, we get $E=4\pi
r_h+3\pi\int_0^{r_h}\frac{\varphi^2r^2}{f(r)}dr$ covering the
Einstein result ($E=4\pi r$) in the appropriate limit of
$\varphi\rightarrow0$ \cite{mis,mis1}. Although, we obtained
$E=4\pi r$ in the $\varphi\rightarrow0$ limit, since we set the
Einstein constant to one ($8\pi G=1$), we have $4\pi=\frac{1}{2G}$
and thus $E=\frac{r}{2G}$ in agreement with previous studies
\cite{ms,mis,mis1}. Substituting Eq.~(\ref{Gtt}) into
Eq.~(\ref{Ein}), one easily finds

\begin{equation}\label{p1}
p=\frac{1}{r^2}(rf^\prime(r)+f(r)-1)+\frac{3}{4f(r)}\varphi^2,
\end{equation}

\noindent which finally leads to
\begin{equation}\label{P2}
pdV=4\pi\Big(rf^\prime(r)+f(r)-1+\frac{3r^2}{4f(r)}\varphi^2\Big)dr,
\end{equation}

\noindent where we used $dV=4\pi r^2dr$. From Eqs.~(\ref{fe})
and~(\ref{P2}), one gets
$\frac{dE}{dr_h}=4\pi+3\pi\textmd{limit}(\frac{\varphi^2r^2}{f(r)})_{r\rightarrow
r_h}$, and
$p\frac{dV}{dr_h}=4\pi(r_hf^\prime(r_h)-1+\textmd{limit}(\frac{3r^2}{4f(r)}\varphi^2)_{r\rightarrow
r_h})$, respectively. It means that, in order to have meaningful
expressions for $\frac{dE}{dr_h}$ and $p\frac{dV}{dr_h}$, the
$\textmd{limit}(\frac{\varphi^2r^2}{f(r)})_{r\rightarrow r_h}$
term should not diverge. Now, using these results as well as the
first law of thermodynamics ($TdS_A=pdV+dE$), we obtain

\begin{equation}
T\frac{dS_A}{dr_h}=6\pi\textmd{limit}(\frac{\varphi^2r^2}{f(r)})_{r\rightarrow
r_h}+4\pi r_hf^\prime(r_h),
\end{equation}

\noindent and therefore

\begin{equation}
T\frac{dS_A}{dr_h}=\frac{f^\prime(r_h)}{4\pi}(\frac{dS_B}{dr_h}+24\pi^2\textmd{limit}(\frac{\varphi^2r^2}{f(r)f^\prime(r)})_{r\rightarrow
r_h}),
\end{equation}

\noindent where $S_B=2\pi A=8\pi^2r_h^2$ is the Bekenstein
entropy. Again, since $8\pi G=c=\hbar=1$, it is easy to obtain the
familiar form of the Bekenstein entropy as $S_B=2\pi
A=\frac{A}{4G}$. Moreover, because we have
$T\equiv\frac{f^\prime(r_h)}{4\pi}$, we finally reach at

\begin{equation}\label{diffentropy}
S_A=S_B+24\pi^2\int\frac{\varphi^2r_h^2}{f(r_h)f^\prime(r_h)}dr_h,
\end{equation}

\noindent for the horizon entropy, covering the Einstein result at
the appropriate limit of $\varphi\rightarrow0$. In order to have a
well-defined entropy, the second term of RHS of this equation
should not be divergent meaning that, at the $r\rightarrow r_h$
limit, the zero-zero component of Lyra time-like displacement
vector field should meet the
$\frac{\varphi}{f(r)}\nrightarrow\infty$ condition.

Finally, let us study the case of $\varphi=l\sqrt{f(r)}$, where
$l$ is arbitrary. Since metric is a spherically symmetric static
metric and thus the Einstein tensor is static, for spherically
symmetric static energy-momentum tensors, $l$ may be a constant or
a function of $r$ and therefore

\begin{equation}
S_A=S_B+24\pi^2\int\frac{l^2r_h^2}{f^\prime(r_h)}dr_h,
\end{equation}

\noindent leading to

\begin{equation}
S_A^d=2\pi A(1-\frac{3l^2}{4\Lambda})
\end{equation}

\noindent and

\begin{equation}
S_A^c=2\pi A[1+\frac{3l^2A^{\frac{3}{2}}}{80m\pi^{\frac{3}{2}}}],
\end{equation}

\noindent for $f(r)=1-\Lambda r^2$
and $f(r)=1-\frac{2m}{r}$ (Schwarzschild),
respectively. In obtaining the above results, we assumed that $l$
is constant. It is obvious that, for real values of $l$, $S_A^c>0$
and we have $S_A^d\geq0$ if and only if $3l^2\leq4\Lambda$. In
addition, it is also easy to verify that, for real values of $l$,
entropy is always positive and non-zero for an anti de-Sitter
universe, where $\Lambda<0$.

\section{Summary}

Throughout this work, applying the unified first law of
thermodynamics to the apparent horizon of FRW universe and using
the Clausius relation as well as the Cai-Kim temperature, we could
obtain a relation for the effects of Lyra displacement vector on
the entropy of apparent horizon. The validity of second and
generalized second laws of thermodynamics have also been studied.

Moreover, considering the Einstein field equations in the Lyra
manifold for a spherically symmetric static metric together with
the first law of thermodynamics, an expression for the entropy of
the static event horizon has also been obtained. Finally, the
horizon entropy for (anti)de-Sitter and Schwarzschild black holes
in the Lyra manifold was also investigated.

\section*{Conflict of Interests}
The authors declare that there is no conflict of interest regarding
the publication of this paper.
\acknowledgments{We are so grateful to the anonymous referees for
their valuable comments. The work of H. Moradpour has been supported
financially by Research Institute for Astronomy \& Astrophysics of
Maragha (RIAAM) under project No.1/4717-170.}

\end{document}